# The evolution of Weyl nodes in Ni doped thallium niobate pyrochlore Tl$_{2-x}$Ni$_x$Nb$_2$O$_7$

Yuefang Hu[a,b], Changming Yue[c], Danwen Yuan[a,b], Jiacheng Gao[d,e], Zhigao Huang[a,b], Zhong Fang[d,e], Chen Fang[d,e], Hongming Weng[d,e,f,*], Wei Zhang[a,b,*]

[a] *Fujian Provincial Key Laboratory of Quantum Manipulation and New Energy Materials, College of Physics and Energy, Fujian Normal University, Fuzhou 350117, China*
[b] *Fujian Provincial Collaborative Innovation Center for Advanced High-Field Superconducting Materials and Engineering, Fuzhou 350117, China*
[c] *Department of Physics, University of Fribourg, 1700 Fribourg, Switzerland*
[d] *Beijing National Laboratory for Condensed Matter Physics, Institute of Physics, Chinese Academy of Sciences, Beijing 100190, China*
[e] *University of Chinese Academy of Sciences, Beijing 100049, China*
[f] *Songshan Lake Materials Laboratory, Dongguan, Guangdong 523808, China*

\* *Correspondence authors. Email:* hmweng@iphy.ac.cn, zhangw721@163.com

**Abstract** Magnetic Weyl semimetal (WSM) is of great importance both for fundamental physics and potential applications due to its spontaneous magnetism, robust band topology, and enhanced Berry curvature. It possesses many unique quantum effects, including large intrinsic anomalous Hall effect, Fermi arcs, and chiral anomaly. In this work, using *ab initio* calculations, we propose that Ni doped pyrochlore Tl$_2$Nb$_2$O$_7$ is a magnetic WSM caused by exchange field splitting on bands around its quadratic band crossing point. The exchange field tuned by Ni 3d on-site Coulomb interaction parameter *U* drives the evolution of Weyl nodes and the resulting topological phase transition. Since Weyl nodes can exist at generic point in Brillouin zone and are hard to be exactly identified, the creation and annihilation of them, i.e., the change in their number, chirality and distribution, have been consistently confirmed with a combined theoretical approach, which employs parity criterion, symmetry indicator analysis and the Wilson loop of Wannier center. We find that Weyl nodes remain in a quite large range of *U* and are close to Fermi level, which makes the experimental observation very possible. We think this method and our proposal of magnetic WSM will be useful in finding more WSMs and adding our understanding on the topological phase transition.

## 1. Introduction

Quantum anomalous Hall effect (QAHE) is a significant transport phenomenon, which has been theoretically proposed [1] and experimentally observed [2] in transition-metal doped thin film of topological insulators of Bi$_2$Se$_3$ family [3,4]. Because of its importance in the fundamental research and potential technical applications, the pursue of QAHE in other topological matters is of great interests [5,6]. It has been proposed that the thin film of magnetic Weyl semimetal (WSM) can realize QAHE even with high Chern number [7,8] and this has been well demonstrated recently in ferromagnetic MnBi$_2$Te$_4$ [9]. Generally, WSMs can be achieved from either non-magnetic WSMs without inversion symmetry (IS), or magnetic WSMs with time-reversal symmetry (TRS) broken. The representative IS-breaking WSMs are those of

TaAs family [10]. These are the first experimentally available materials in WSM phase. Many exotic features of WSM have been studied in them, such as the observation of bulk Weyl nodes, surface Fermi arcs, chiral anomaly, and nonlinear optical effects [11-14]. On the other hand, the magnetic WSMs have been proposed theoretically earlier than the nonmagnetic ones, such as $Y_2Ir_2O_7$ [15], $HgCr_2Se_4$ [7] and some Heusler compounds [16,17]. It's not until very recently that several magnetic WSMs have been confirmed. They have quite large anomalous Hall effect (AHE), such as $Co_3Sn_2S_2$ [18-22], $Co_2MnGa$ [23] and $Mn_3Sn$ [24,25], due to the nearly divergent Berry curvature near the Weyl nodes. In addition to hosting AHE or QAHE, magnetic WSMs can offer another knob to tune the Weyl nodes through controlling their magnetization. The evolution of Weyl nodes is quite interesting since it is closely related with the creation and annihilation of Weyl nodes, topological phase transition, AHE, and magneto-optical properties. Therefore, it's in urgent need to explore more new intrinsic magnetic WSMs [26,27], which can not only provide a platform for studying the interaction among magnetism, band topology and electron correlation, but also guide the directions for uncovering remarkable responses for potential applications.

In this work, we propose Ni doped $Tl_{2-x}Ni_xNb_2O_7$ (x = 0.0625) to be an intrinsic magnetic WSM. The calculated Weyl nodes have been demonstrated to evolve with Coulomb interaction parameter $U$ on Ni site, which tunes the magnetization and the exchange field splitting on the bands around the quadratic band crossing point (QBCP) in parent compound $Tl_2Nb_2O_7$. Thallium niobate $Tl_2Nb_2O_7$ possesses pyrochlore structure, whose high-quality single crystal has been well fabricated since 1960s [28,29]. Especially, the system remains in pyrochlore structure stably as a large amount of Tl (up to 75%) is replaced by Ag/Pb (or some alkalin-earth metals), confirmed by the X-ray analysis [30]. On the other hand, the oxygen-deficient sample $Tl_2Nb_2O_{6+y}$ is available with y being continuously varied between 0 and 1. $Tl_2Nb_2O_{6+y}$ has been proposed to host various topological states, including Dirac semimetal, triply degenerate nodal point semimetal and topological insulator via strain and/or oxygen deficiency in our previous study [31].

Since the Weyl nodes can exist at generic momentum points, locating the exact positions of them is quite a challenge. We have employed three methods, namely the parity criterion [32], symmetry indicator analysis [33,34] and Wilson loop of Wannier center, to consistently and collaboratively identify their number, chirality and distribution. Within our calculations, $Tl_{2-x}Ni_xNb_2O_7$ is a magnetic WSM with inversion symmetry preserved and the number of Weyl nodes changes with $U$. The creation and annihilation of Weyl nodes have been carefully studied and the resulted topological phase transitions are shown. Thus, the current system with inversion symmetry provides a playground for investigating the topological properties and phase transitions in magnetic WSM.

## 2. Calculation methods

We have employed Vienna *ab initio* simulation package (VASP) [35,36] to perform first-principles calculations within density functional theory (DFT). The generalized gradient approximation (GGA) in the form of the Perdew, Burke, and Ernzerhof is applied for treating the exchange-correlation functional [37]. The kinetic energy cutoff of plane wave expansion is 300 eV. In self-consistent electronic structure calculations, we use 6 × 6 × 6 k-point grids. The initial spin-polarization of Ni ion along *z*-axis is adopted when spin-orbit coupling (SOC) is included. In order to deal with the electron-electron correlation interaction among the localized 3d electrons on Ni site, Hubbard $U$ correction is implemented within the GGA+$U$ scheme [38] and the value of *U* is changed as a parameter in the range from 0 to 4 eV. Atomic orbital like Wannier functions have been generated [39,40] for Ni 3d, Tl 6s, Nb 4d, and O 2p. They have been passed to WannierTools [41] to calculate the topological properties of Weyl nodes.

## 3. Results and discussion

*3.1. Crystal structure*

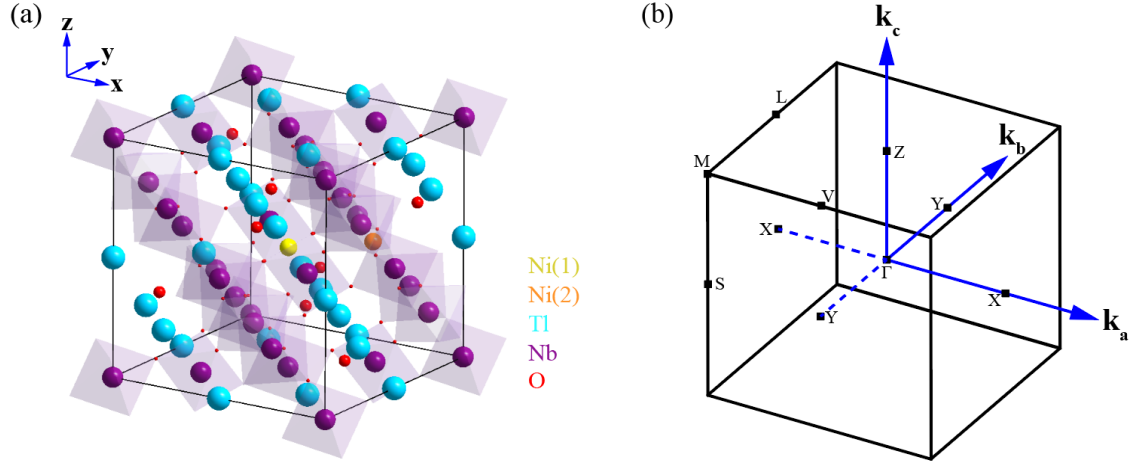

**Fig. 1.** (Color online) The crystal structure of $Tl_{2-x}Ni_xNb_2O_7$ (x = 0.0625). (a) Primitive unit cell of the doped system. Two nonequivalent Tl sites can be replaced with Ni [Ni(1) and Ni(2)]. Ni(1) indicates the body center position, while Ni(2) indicates the face center position in the primitive cell of $Tl_2Nb_2O_7$. (b) The related first Brillouin zone.

$Tl_2Nb_2O_7$ crystallizes in a pyrochlore structure with space group *Fd-3m* (No. 227; Z = 2) [28,29]. Tl, Nb atoms are located at 16d and 16c Wyckoff positions, respectively, and there are two sites for oxygen atoms, namely O and O' at 48f and 8b. O' is at the center of Tl tetrahedron and can form deficiency with its concentration continuously varying between 0 and 1 [31]. This deficiency will cause the shift of Tl and break inversion symmetry, which induces topological phase transitions from a parabolic band touching nodal point semimetal phase to a triply degenerate nodal point semimetal phase, and to a trivial narrow gap semiconductor phase [31]. All these states are possibly to be tuned to the magnetic topological Weyl semimetal if TRS is broken with proper exchange field. This scheme can be stimulated by inspecting the Cr-doped thin film of topological insulator $Bi_2Se_3$ to realize QAHE [1].

Since the $NbO_6$ forms stable octahedron, the tetrahedron composed of Tl and O' serves as the knob to control the electronic structures of this system by oxygen deficiency and Tl valence changing from $Tl^+$ to $Tl^{3+}$ [31]. On considering this and in order to break TRS, we dope magnetic ions by replacing Tl. Experimentally, it can keep pyrochlore structure even if Tl is replaced by Ag or Pb up to 75% [30]. We choose transition metal element Ni as an example to discuss in the following, namely $Tl_{2-x}Ni_xNb_2O_7$ with the concentration of 6.25%.

As shown in the Fig. 1(a), the unit cell of 6.25% Ni-doped $Tl_2Nb_2O_7$ has 88 atoms (Z = 8) in total. The doped system has two different configurations, depending on the substitution of Ni atom for Tl atom at the body center [Ni(1)] or face center [Ni(2)] position in the primitive cell of $Tl_2Nb_2O_7$. The total energy of the Ni(1)-doped configuration is lower than that of the Ni(2)-doped one by ~0.211 meV/Å$^3$. They have the similar topological properties. Thus, in the following, we will focus on the discussion about the Ni(1)-doped case throughout the paper, while the results of Ni(2)-doped case are shown in the Supplementary materials (online). After full structural relaxation, the crystal structure of the doped system is in space group $C_{2/m}$ (No. 12), and the related first Brillouin zone (BZ) is shown in Fig. 1(b). The optimized lattice constants are $a = b = 10.702$ Å, $c = 10.700$ Å. In the calculation with SOC, the magnetic moment is along $z$-axis and the magnetic space group is $C_{2'/m'}$ (No. 12.62). When the symmetry indicator of inversion symmetry $Z_4$ is odd, $C_2*T$ operation limits odd number of pairing Weyl nodes in the plane conserving it. $C_2$ is the two-fold rotation operation and $T$ is time reversal operation [34].

*3.2. Topological properties*

It is known that before doping, $Tl_2Nb_2O_7$ is a zero-gap semimetal with QBCP at Γ point, which is composed of hybridization between oxygen 2p and Tl 5d-$t_{2g}$ orbitals. The hybridization is prominent since the effective SOC in this triply degenerate state is negative. Starting from this state, Dirac semimetal and topological insulator states can be obtained by applying proper strain. This is also a good starting point to obtain magnetic topological states by introducing time-reversal breaking magnetic effect. One proper way is magnetic ion doping, which has successfully been used to generate quantized AHE in two dimensional (2D) systems [1]. The magnetic topological states in three-dimensional (3D) solids might include magnetic Weyl semimetal, axion insulator and other magnetic crystalline insulators with extra spatial symmetry [34]. After Ni doping, the local magnetic momentum on Ni ions will generate exchange splitting on the bands around QBCP through hybridization. The exchange splitting will be crucial to tune the magnetic topological states as we will discuss below.

The band structures of ferromagnetic $Tl_{2-x}Ni_xNb_2O_7$ (x = 0.0625) calculated with different $U$ value from 0 to 4 eV are shown in Fig. 2. All these band structures are close to semimetal features with several band crossing points near the Fermi level. Since inversion symmetry is preserved, the wavefunction at time-reversal momenta (TRIM) is also the eigenstate of inversion operation and the parities of the occupied bands and some of the unoccupied ones have been calculated. The number of occupied odd parity bands at eight TRIM for different $U$ values are listed in Fig. 2(f). Some even and odd parity bands around Fermi level are labeled with + and - sign, respectively. It is noted that as $U$ increases, the number of odd parity bands changes at TRIM Z, S, L and V. Since the red bands have a quite large gap, these changes come from the band inversion among the blue bands as shown in Fig. 2. Obviously, these changes will cause topological phase transition.

When SOC is further considered and the magnetization of the local moment on Ni is assumed to along the $z$-axis, a similar band dispersion and the same parity configuration are obtained since inversion symmetry is still preserved. The SOC results are exhibited in Fig. S1 (online). In order to identify the topological phases of $Tl_{2-x}Ni_xNb_2O_7$ (x = 0.0625), we calculate the stable and weak topological indices $Z_4$ and $Z_{2i}$ ($i = a, b, c$) determined by parity eigenvalues at eight TRIM [33,34]. They are defined as:

$$Z_4 = \sum_K n_K^- \bmod 4, \text{ and } Z_{2i} = \sum_{K, K_i = \pi} n_K^- \bmod 2$$

where the $K$ sums over all the TRIM and $n_K^-$ stands for the number of occupied states with odd parity eigenvalues at momentum $K$. For $U$ = 0 eV, the topological indices ($Z_4$; $Z_{2a}$, $Z_{2b}$, $Z_{2c}$) is (0; 1,1,0), which might be a 3D QAHE state or a WSM with even number of Weyl nodes in half BZ. For $U$ = 1 and 2 eV, $Z_4$ = 1 indicates they are WSMs with odd number of Weyl nodes in half BZ, to be more specific, in half of the $C_2*T$ invariant plane. For $U$ = 3 and 4 eV, ($Z_4$; $Z_{2a}$, $Z_{2b}$, $Z_{2c}$) is (2; 1,1,1), corresponding to a 3D QAHE state or a WSM with even number of Weyl nodes in half BZ. These imply that there are two topological phase transitions occurring as $U$ changes from 0 eV to 1 eV and from 2 eV to 3 eV. According to the study of topological phase transition [42,43], there must be creation and annihilation of Weyl nodes accompanied and these can happen at TRIM like Z, S, L, V, where the number of occupied odd parity bands changes. The number of occupied odd parity bands can be odd or even at each TRIM as shown in Fig. 2f.

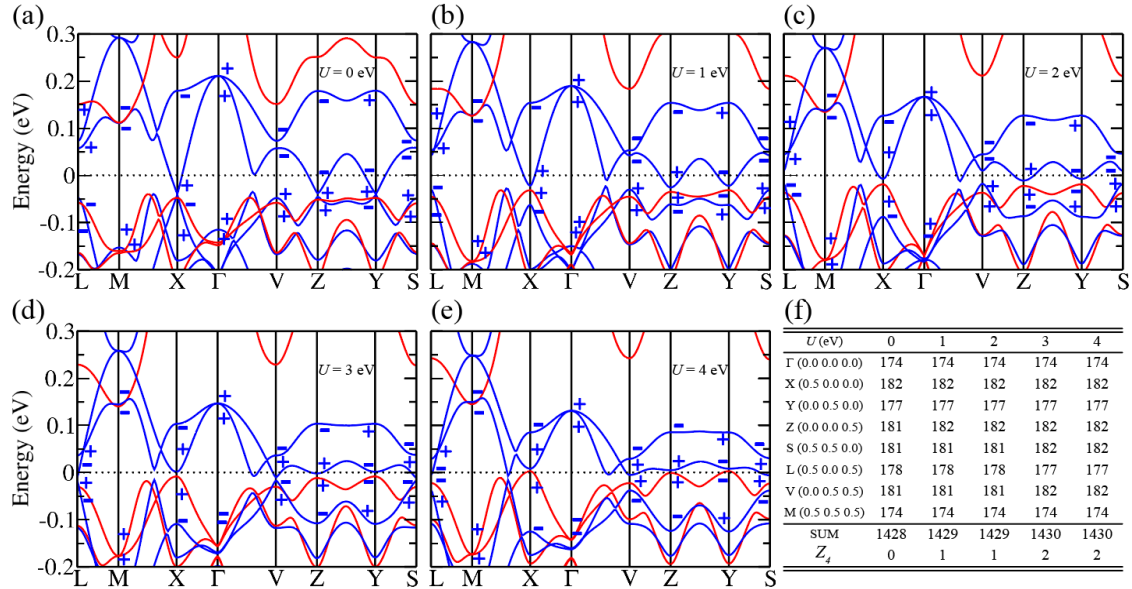

**Fig. 2.** (Color online) Band structures for the ferromagnetic $Tl_{2-x}Ni_xNb_2O_7$ (x = 0.0625) calculated from GGA+$U$ method without SOC for (a) $U$ = 0 eV, (b) $U$ = 1 eV, (c) $U$ = 2 eV, (d) $U$ = 3 eV, (e) $U$ = 4 eV. The blue and red bands represent the spin up and spin down states, respectively. The parities of top two occupied bands and bottom two unfilled bands at TRIM are marked as + (-) beside the bands. (f) The number of occupied odd parity bands and $Z_4$ at eight TRIM for each $U$.

*3.3. Evolution of Weyl fermions*

Motivated by the above analysis, we further investigate all the generic k-points in the 3D BZ in order to search for the locations of Weyl nodes with SOC included, which should depend on the on-site electron-electron correlation parameter $U$. The position, chirality, energy level, and the number of Weyl nodes in each case are summarized in Table S1 and Fig. S1(f) (online). All these Weyl nodes have the chirality of +1 or -1 and are found to be away from Fermi level within 35.6 meV, which might be good for experimental detection and appearance of topological effects such as Fermi arcs and chiral anomaly. There are 8, 6, 6 Weyl nodes in half BZ as $U$ =

0, 3, 4 eV respectively, while there are 5 ones for $U = 1$ and 2 eV. Thus, the odd or even number of the Weyl nodes agrees well with the above $Z_4$ analysis.

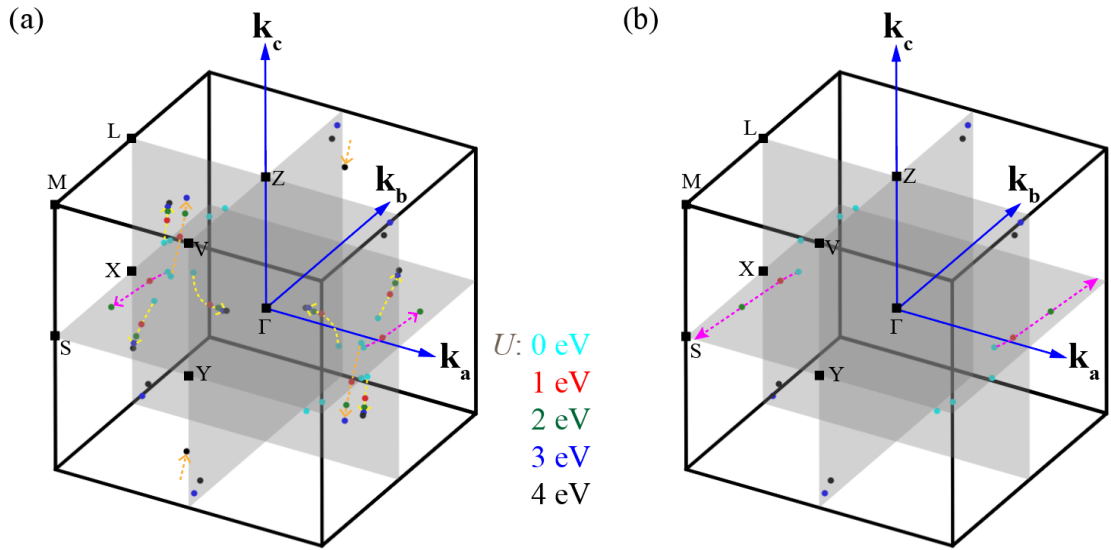

**Fig. 3.** (Color online) Evolution of the Weyl points in the 3D BZ with the electron-electron correlation $U$ when SOC is considered. Dots in various color represent the positions of Weyl nodes calculated with different values of $U$. (a) Distribution of the Weyl nodes, whose chirality are all +1 or -1. The dotted arrows trace the motion of the Weyl nodes in the direction of increasing $U$. The yellow arrows give out a complete movement of the Weyl nodes in the first BZ as $U$ increases from 0 to 4 eV. Meanwhile, the Weyl nodes indicated by the orange arrows actually across the boundary of the first BZ and back in from the other side of BZ by shifting one reciprocal lattice of $k_c$. (b) The emergence or annihilation of the Weyl points as $U$ changes. Three pairs of cyan dots annihilate at Z points as $U$ increases from 0 eV to 1 eV. A pair of Weyl points denoted by purple arrows approaches each other and annihilates at S point, while two pairs of blue dots emerge from V and L points, when $U$ increases from 2 eV to 3 eV.

The distribution of all Weyl nodes satisfies the inversion symmetry, as shown in the Fig. 3(a). The two Weyl points related by the inversion symmetry are in the same energy, but with opposite chirality. When $U$ increases from 0 eV to 1 eV, three of the eight pairs of Weyl nodes in cyan dots can shift to and annihilate at Z point, which is consistent with the number of odd parity bands changing only at Z. This leads to a topological phase transition to WSM state at $U = 1$ eV. The other five pairs of Weyl nodes from $U = 0$ eV shift in a path as indicated by five pairs of arrows. As $U$ further increases to 2eV, the parity configuration at all eight TRIM is not changed and there is no topological phase transition, and the five pairs of Weyl nodes are shifted further to green dots as indicated by the arrows. From $U = 2$ eV to 3 eV, the five pairs of Weyl nodes move along the arrows and one pair of them annihilate at S point, as illustrated by the purple arrows in Fig. 3 (a) and (b). Meanwhile, there are two pairs of Weyl nodes newly created around V and L. One pair of them is very near to $k_a = 0$ plane, while the other pair is close to $k_b = 0$ plane, as shown by the blue dots near V and L points in Fig. 3(b). Therefore, there are totally six pairs of Weyl nodes at $U = 3$ eV. During the phase transitions, the moving, creation and annihilation of the Weyl nodes are consistent with the changes in the number of odd parity bands at Z, S, L and V. When $U$ increases to 4 eV, the blue dots (Weyl nodes in $U = 3$eV) move to the place denoted by black dots. The blue dots indicated by the orange arrows move out of the first BZ and back in from the other side of BZ by shifting one reciprocal lattice of $k_c$. The increasing of $U$ drives the Weyl nodes to the positions at lower energies and most of them tend to leave the center of the BZ. When $U \geq 3$ eV, the positions of Weyl nodes are all below the Fermi level, which facilitates the angle-resolved photoemission spectroscopy (ARPES) measurement.

*3.4. Topological charge of the Weyl points*

To further shed light on the phase transition mechanism and the distribution of chiral Weyl nodes with their topological charges, we calculate the Wilson loops of occupied states (those below and including the low branch of Weyl nodes) for $k_i = 0$ or $k_i = \pi$ plane (*i=a, b or c*). As shown in the Fig. 4(a), when $U = 2$ eV, the Wilson loop calculations suggest that the Chern number, defined as the winding number of the avarage Wannier center along one of the periodic directions, is $C = 0$ at $k_i = 0$ plane, and $C = 1$ or $C = -1$ at $k_i = \pi$ plane. This is consistent with the parity configuration shown in Fig. 2(f) and Fig. S2(b) (online), since it is noted that the parity criterion can only determine the Chern number value difference modulo 2. The difference of the Chern number between $k_i = 0$ and $k_i = \pi$ is 1, which indicates there must be odd number of Weyl nodes in the half BZ between these two planes and the net topological charge of these Weyl nodes is 1. This can be checked by projecting the Weyl nodes onto $k_a$-$k_b$, $k_a$-$k_c$ or $k_b$-$k_c$ plane, respectively, as shown in Fig. 4(c) by the green dots. The topological charge of each node is identified by + or - sign. It is easy to see that there are five Weyl nodes in half BZ and the net charge value of them is 1.

Similarly, when $U = 3$ eV, the Wilson loops in Fig. 4(b) indicate that all the three pairs of $k_i = 0$ and $k_i = \pi$ planes have absolute Chern number of 1, which is also consistent with the parity configuration in Fig. 2(f) and Fig. S2(c) (online). The Chern number difference between $k_c = 0$ and $k_c = \pi$ planes is two, while that between other two pairs is zero. This is consistent with the value of $Z_4 = 2$ since $Z_4$ mod 2 indicates the parity of Chern number difference between these plane pairs. Therefore, there are even number of Weyl nodes in the half BZ between these three pairs of planes, but the net charge should be two for that between $k_c = 0$ and $k_c = \pi$ planes and zero for the other two. In order to check these arguments, we project the positions of Weyl points obtained in $U = 3$ eV onto three planes of $k_a$, $k_b$, and $k_c = 0$. As shown by the blue dots in Fig. 4(c). The topological charge of each node is identified by + or - sign. From this plot, we can see that there are six Weyl nodes in all half BZ. The absolute value of the net topological charge in the half BZ along $k_a$ or $k_b$ direction is zero, while it is two along $k_c$ direction.

Moreover, comparing the positions of Weyl nodes in $U = 2$ eV with those in $U = 3$ eV, we can see that on-site Coloumb interaction parameter *U* drives the movement of Weyl nodes. As displayed in the left panel of Fig. 4(c), a pair of green dots (for $U = 2$ eV) with opposite chirality approaches the low-left corner and top-right corner, where S is projected on, respectively, and they annihilate when *U* increases to 3 eV. This is consistent with the change in number of odd parity bands at S from 181 to 182, as displayed in Fig. 2(f). Meanwhile, two pairs of Weyl nodes are created. As shown in Fig. 4(c), one pair is the blue dots (for $U = 3$ eV) close to $k_a = 0$ line and the projection of V. The other pair is close to $k_b = 0$ line and the projection of L. Obviously, these two pairs of Weyl nodes come from the band-inversion-induced change in number of odd parity bands at TRIM V and L. The evolution of Weyl nodes also agrees well with that shown in Fig. 3(b).

Since the magnetic space group is $C_{2'/m'}$ (No. 12.62), there is odd number of pairing Weyl nodes in the plane conserving $C_2*T$ [34] when $Z_4$ is odd. In Fig. 4c, the diagonal red line is the projection of the plane invariant under $C_2*T$. It is noted that there are three pairs of Weyl nodes in the plane for $U = 2$ eV and two pairs for $U = 3$ eV, which is consistent with the limitation on the distribution of Weyl nodes applied by $Z_4$ and $C_2*T$. As discussed above, one pair of Weyl nodes in $U = 2$ eV moves towards TRIM S and annihilates there by band inversion, indicated by the change in the number of occupied odd parity bands.

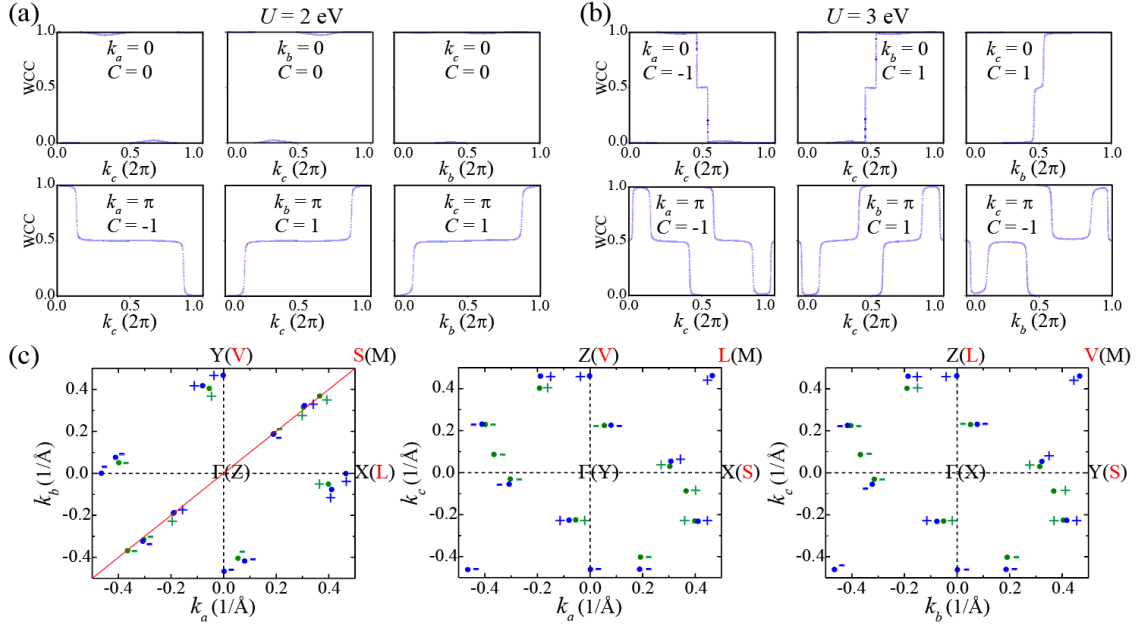

**Fig. 4.** (Color online) The evolution of the average position of Wannier centers in the $k_a$ ($k_b$, $k_c$) = 0, π plane, as (a) $U$ = 2 eV and (b) $U$ = 3 eV. (c) The distribution of the Weyl nodes in the $k_c$ ($k_b$, $k_a$) = 0 plane. The green and blue dots denote the projected positions of Weyl nodes as $U$ = 2 eV and $U$ = 3 eV, respectively. The + (-) beside each Weyl node indicates the topological charge +1 (-1). The TRIM in red indicates where the creation and annihilation of Weyl nodes happen.

## 4. Conclusion

In brief, motivated by the QBCP in pyrochlore $Tl_2Nb_2O_7$, we have proposed that the diluted magnetic ion doping can induce TRS-breaking WSM phase in it through the exchange field splitting effect. Using the first-principles calculations, we have shown that the Ni doped $Tl_{2-x}Ni_xNb_2O_7$ (x = 0.0625) can have different topological phases with Weyl nodes close to Fermi level, which is driven by the on-site Coulomb interaction of Ni 3d electrons parameterized by $U$. The topological phase transitions happen due to the creation and annihilation of Weyl nodes accompanied by band inversions at TRIM. Thanks to the inversion symmetry preserved in the system, we have combined the advantages of three methods, namely the parity criterion, symmetry indicator analysis, and Wilson loop of Wannier center, to identify the number of Weyl nodes, their chirality and distributions in different $U$ value cases. There are eight pairs of Weyl nodes when $U$ = 0 eV. Three of them annihilate when $U$ increases to 1 eV with a band inversion at Z and it becomes a topologically robust WSM. When $U$ increases to 3 eV, one pair of Weyl nodes annihilates at S and other two pairs emerge around L and V, respectively, accompanied with band inversions at these TRIM. Throughout the range of $U$ from 0 to 4 eV, the Weyl points at generic points remain existing, and this will facilitate the experimental observation of them through ARPES technique. The combined method we described here to identify Weyl nodes will also be very useful and applicable to all magnetic topological materials with inversion symmetry.

## Conflict of interest

The authors declare that they have no conflict of interest.

## Acknowledgments

This work was supported by the National Natural Science Foundation of China (11974076, 11925408, 11921004 and 12188101), the Key Project of Natural Science Foundation of Fujian Province (2021J02012), the Ministry of Science and Technology of China (2018YFA0305700), the Chinese Academy of Sciences (XDB33000000), the K. C. Wong Education Foundation (GJTD-2018-01). Changming Yue was supported by the Swiss National Science Foundation (200021-196966).

**Author contributions**

Yuefang Hu performed the DFT calculations and Changming Yue performed the parity calculations. Danwen Yuan, Jiacheng Gao, Zhigao Huang, Chen Fang and Zhong Fang contributed to analyzing the data. Yuefang Hu, Wei Zhang and Hongming Weng wrote the manuscript. Wei Zhang and Hongming Weng supervised the project. All authors participated in the discussions.

**Appendix A. Supplementary materials**

Supplementary materials to this article can be found online at

**References:**

[1] Yu R, Zhang W, Zhang H, et al. Quantized anomalous Hall effect in magnetic topological insulators. Science 2010;329:61-64.
[2] Chang C, Zhang J, Feng X, et al. Experimental observation of the quantum anomalous Hall effect in a magnetic topological insulator. Science 2013;340:167-170.
[3] Zhang H, Liu C, Qi X, et al. Topological insulators in $Bi_2Se_3$, $Bi_2Te_3$ and $Sb_2Te_3$ with a single Dirac cone on the surface. Nat Phys 2009;5:438-442.
[4] Zhang W, Yu R, Zhang H, et al. First-principles studies of the three-dimensional strong topological insulators $Bi_2Te_3$, $Bi_2Se_3$ and $Sb_2Te_3$. New J Phys 2010;12:065013.
[5] Feng X, Jiang K, Wang Z, et al. Chiral flux phase in the Kagome superconductor $AV_3Sb_5$. Sci Bull 2021;66:1384-1388.
[6] Xu L, Mao Y, Wang H, et al. Persistent surface states with diminishing gap in $MnBi_2Te_4$/$Bi_2Te_3$ superlattice antiferromagnetic topological insulator. Sci Bull 2020;65:2086-2093.
[7] Xu G, Weng H, Wang Z, et al. Chern semimetal and the quantized anomalous Hall effect in $HgCr_2Se_4$. Phys Rev Lett 2011;107:186806.
[8] Weng H, Yu R, Hu X, et al. Quantum anomalous Hall effect and related topological electronic states. Adv Phys 2015;64:227-282.
[9] He K. $MnBi_2Te_4$-family intrinsic magnetic topological materials. npj Quantum Mater 2020;5:90.
[10] Weng H, Fang C, Fang Z, et al. Weyl semimetal phase in noncentrosymmetric transition-metal monophosphides. Phys Rev X 2015;5:011029.
[11] Wu L, Patankar S, Morimoto T, et al. Giant anisotropic nonlinear optical response in transition metal monopnictide Weyl semimetals. Nat Phys 2017;13:350-355.
[12] Osterhoudt GB, Diebel LK, Gray MJ, et al. Colossal mid-infrared bulk photovoltaic effect in a type-I Weyl semimetal. Nat Mater 2019;18:471-475.
[13] Ma J, Gu Q, Liu Y, et al. Nonlinear photoresponse of type-II Weyl semimetals. Nat Mater 2019;18:476-481.
[14] Gao Y, Kaushik S, Philip EJ, et al. Chiral terahertz wave emission from the Weyl semimetal TaAs. Nat Commun 2020;11:720.
[15] Wan X, Turner AM, Vishwanath A, et al. Topological semimetal and Fermi-arc surface states in the electronic structure of pyrochlore iridates. Phys Rev B 2011;83:205101.
[16] Manna K, Sun Y, Muechler L, et al. Heusler, Weyl and Berry. Nat Rev Mater 2018;3:244-256.
[17] Wang Z, Vergniory MG, Kushwaha S, et al. Time-reversal-breaking Weyl fermions in magnetic Heusler alloys. Phys Rev Lett 2016;117:236401.
[18] Wang Q, Xu Y, Lou R, et al. Large intrinsic anomalous Hall effect in half-metallic ferromagnet $Co_3Sn_2S_2$ with magnetic Weyl fermions. Nat Commun 2018;9:3681.
[19] Liu E, Sun Y, Kumar N, et al. Giant anomalous Hall effect in a ferromagnetic kagome-lattice semimetal. Nat Phys 2018;14:1125-1131.

[20] Liu DF, Liang AJ, Liu EK, et al. Magnetic Weyl semimetal phase in a Kagomé crystal. Science 2019;365:1282-1285.
[21] Morali N, Batabyal R, Nag PK, et al. Fermi-arc diversity on surface terminations of the magnetic Weyl semimetal $Co_3Sn_2S_2$. Science 2019;365:1286-1291.
[22] Weng H. Magnetic Weyl semimetal finally confirmed. Sci China Phys Mech Astron 2019;62:127031.
[23] Belopolski I, Manna K, Sanchez DS, et al. Discovery of topological Weyl fermion lines and drumhead surface states in a room temperature magnet. Science 2019;365:1278-1281.
[24] Yang H, Sun Y, Zhang Y, et al. Topological Weyl semimetals in the chiral antiferromagnetic materials $Mn_3Ge$ and $Mn_3Sn$. New J Phys 2017;19:015008.
[25] Matsuda T, Kanda N, Higo T, et al. Room-temperature terahertz anomalous Hall effect in Weyl antiferromagnet $Mn_3Sn$ thin films. Nat Commun 2020;11:909.
[26] Lu Y, Wang B, Liu X. Ideal Weyl semimetal with 3D spin-orbit coupled ultracold quantum gas. Sci Bull 2020;65:2080-2085.
[27] Gao J, Qian Y, Nie S, et al. High-throughput screening for Weyl semimetals with $S_4$ symmetry. Sci Bull 2021;66:667-675.
[28] Uma S, Kodialam S, Yokochi A, et al. Structure and properties of $Tl_2Nb_2O_{6+x}$ phases with the pyrochlore structure. J Solid State Chem 2000;155:225-228.
[29] Fourquet JL, Duroy H, Lacorre P. $Tl_2Nb_2O_{6+x}$ ($0 \leq x \leq 1$): a continuous cubic pyrochlore type solid solution. J Solid State Chem 1995;114:575-584.
[30] Fourquet JL, Plet F, de Pape R. Proprietes d'echange ionique du pyrochlore $TlNbO_3$ en presence de nitrates fondus ou solides. Mater Res Bull 1975;10:933-939.
[31] Zhang W, Luo K, Chen Z, et al. Topological phases in pyrochlore thallium niobate $Tl_2Nb_2O_{6+x}$. npj Comput Mater 2019;5:1-7.
[32] Fu L, Kane CL. Topological insulators with inversion symmetry. Phys Rev B 2007;76:045302.
[33] Xu Y, Elcoro L, Song Z, et al. High-throughput calculations of magnetic topological materials. Nature 2020;586:702-707.
[34] Peng B, Jiang Y, Fang Z, et al. Topological classification and diagnosis in magnetically ordered electronic materials. arXiv:2102.12645, 2021.
[35] Kresse G, Furthmüller J. Efficiency of ab-initio total energy calculations for metals and semiconductors using a plane-wave basis set. Comput Mater Sci 1996;6:15-50.
[36] Kresse G, Furthmüller J. Efficient iterative schemes for ab initio total-energy calculations using a plane-wave basis set. Phys Rev B 1996;54:11169-11186.
[37] Perdew JP, Burke K, Ernzerhof M. Generalized gradient approximation made simple. Phys Rev Lett 1996;77:3865-3868.
[38] Dudarev SL, Botton GA, Savrasov SY, et al. Electron-energy-loss spectra and the structural stability of nickel oxide: An LSDA + U study. Phys Rev B 1998;57:1505-1509.
[39] Marzari N, Vanderbilt D. Maximally localized generalized Wannier functions for composite energy bands. Phys Rev B 1997;56:12847-12865.
[40] Souza I, Marzari N, Vanderbilt D. Maximally localized Wannier functions for entangled energy bands. Phys Rev B 2001;65:035109.
[41] Wu Q, Zhang S, Song H, et al. WannierTools: An open-source software package for novel topological materials. Comput Phys Commun 2018;224:405-416.
[42] Murakami S. Phase transition between the quantum spin Hall and insulator phases in 3D: emergence of a topological gapless phase. New J Phys 2007;9:356.
[43] Murakami S. Gap closing and universal phase diagrams in topological insulators. Physica E Low Dimens Syst Nanostruct 2011;43:748-754.